\documentclass[10pt]{article}	% essential first line
\usepackage{float}		% this is to place figures where requested!
\usepackage{times}		% this uses fonts which will look nice in PDF format
\usepackage{graphicx}		% needed for the figures
\usepackage{url}
\usepackage{epstopdf}
\usepackage{longtable}
\usepackage{amsmath}
\usepackage{amssymb}
\usepackage{multicol}
\usepackage[font={small}]{caption}
\usepackage{subcaption}
\usepackage{abstract}
\usepackage[procnames]{listings}
\usepackage{tabto}
\usepackage{paralist} % Used for the compactitem environment which makes bullet points with less space between them
\usepackage{xcolor}
\usepackage[document]{ragged2e}
\usepackage{txfonts} %this gives a compact text style
\usepackage{microtype}
\usepackage{enumitem}  % this gives compact lists
\usepackage{comment} 
\usepackage[T1]{fontenc}
\usepackage{wrapfig}
\usepackage{tabularx}

\usepackage{placeins}
\usepackage{caption}
\usepackage[round]{natbib}
\usepackage{bm}
\usepackage{listings}
\usepackage{hyperref}

\newcommand{\aap}{A\&A}

\newcommand{\apjl}{ApJL}

\newcommand{\apjs}{ApJS}

\hypersetup{
    colorlinks,
    linkcolor={blue!70!black},
    citecolor={blue!70!black},
    urlcolor={blue!70!black}
}
\newcommand{\farcs}{\ensuremath{.\!\!^{\prime\prime}}} 
\newcommand{\arcsec}{\ensuremath{^{\prime\prime}}} 
\newcommand{\arcmin}{\ensuremath{^{\prime}}} 

\DeclareGraphicsExtensions{.pdf,.png,.jpeg}

\def\memonr{2023-L1-02}

\title{\vspace{-1.25cm} 
\textcolor{gray}{\textit{L1 Processing and Validation}}\\
\vspace{0.40cm}
\textbf{Absolute Flux Density Calibration of the Greenland Telescope Data for Event Horizon Telescope Observations}
} 

\author{ \renewcommand{\thefootnote}{\arabic{footnote}} 
J. Y. Koay$^{1}$, 
K. Asada$^{1}$, 
S. Matsushita$^{1}$,
C.-Y. Kuo$^{2,1}$,
C.-W. L. Huang$^{1}$,
C. Romero-Ca\~nizales$^{1}$,\\
S. Koyama$^{3,1}$,
J. Park$^{4}$,
W.-P. Lo$^{1}$,
G. Bower$^{1}$, 
M.-T. Chen$^{1}$, 
S.-H. Chang$^{1}$,
C.-C. Chen$^{1}$,
R. Chilson$^{1}$, \\
C. C. Han$^{1}$,
P. T. P. Ho$^{1}$,
Y.-D. Huang$^{1}$, 
M. Inoue$^{1}$,
B. Jeter$^{1}$,
H. Jiang$^{1}$,
P. M. Koch$^{1}$, 
D. Kubo$^{1}$,
C.-T. Li$^{1}$, \\ 
C.-T. Liu$^{1}$, 
K.-Y. Liu$^{1}$,
P. Martin-Cocher$^{1}$,
M. Nakamura$^{5,1}$,
T. J. Norton$^{6}$, 
G. Nystrom$^{1}$,
P. Oshiro$^{1}$, \\
N. Patel$^{6}$, 
U.-L. Pen$^{1}$,
H.-Y Pu$^{7,1}$,
P. A. Raffin$^{1}$, 
R. Rao$^{6}$, 
T. K. Sridharan$^{6}$,
R. Srinivasan$^{6}$,
T.-S Wei$^{1}$
}

\def\affiliations{
\hspace{0.5cm}$^{1}$\textit{Academia Sinica Institute of Astronomy and Astrophysics (ASIAA), No.1, Sec. 4, Roosevelt Rd, Taipei 10617, Taiwan, R.O.C.} \\
\hspace{0.5cm}$^{2}$\textit{Physics Department, National Sun Yat-Sen University, No. 70, Lien-Hai Rd, Kaosiung City 80424, Taiwan, R.O.C} \\
\hspace{0.5cm}$^{3}$\textit{Niigata University, 8050 Ikarashi-nino-cho, Nishi-ku, Niigata 950-2181, Japan} \\
\hspace{0.5cm}$^{4}$\textit{Department of Astronomy and Space Science, Kyung Hee University, 1732, Deogyeong-daero, Giheung-gu, Yongin-si, Gyeonggi-do 17104, Republic of Korea} \\
\hspace{0.5cm}$^{5}$\textit{National Institute of Technology, Hachinohe College, 16-1 Uwanotai, Tamonoki, Hachinohe, Aomori 039-1192, Japan} \\
\hspace{0.5cm}$^{6}$\textit{Center for Astrophysics $|$ Harvard \& Smithsonian, 60 Garden St., Cambridge, MA 02138, USA} \\
\hspace{0.5cm}$^{7}$\textit{National Taiwan Normal University, Taiwan, R.O.C} \\
}

\date{Oct 10, 2022 -- Version 1.0}
\def\memohistory{
\item {Dec 10, 2021 -- Document created}
}
\newif\ifshowhistory  
\begin{document} 

\begin{figure}[!t]
\begin{minipage}[t]{0.49\textwidth}
    \vspace{-0.13\linewidth}{\includegraphics[width=0.6\linewidth]{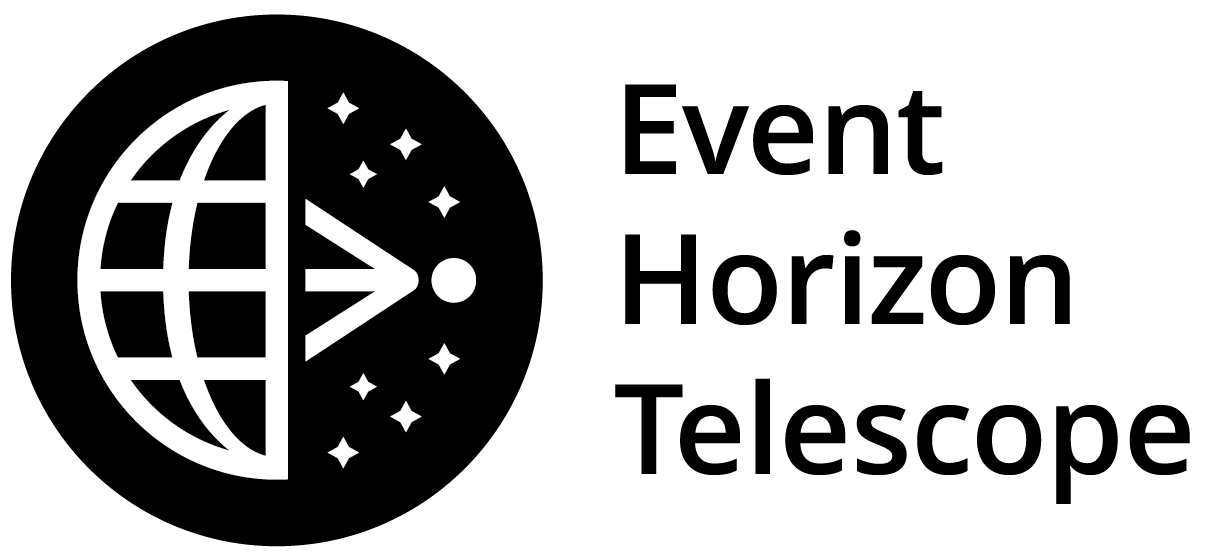}}
    \label{fig:my_label}
\end{minipage}
\begin{minipage}[t]{0.49\textwidth}
\begin{flushright} \large
\textbf{Event Horizon Telescope \\ Memo Series} 
\end{flushright}
\end{minipage}
\end{figure}

\begin{center}
EHT Memo \memonr{}
\end{center}

% Title section: Use the \author, \title, and \date info
{\let\newpage\relax\maketitle}  % Prevent maketitle inserting newpage if possible
\thispagestyle{empty}		% Inhibit the page number on this page

% Add Affiliations
\begin{flushleft}
{\small
\affiliations{}
}
\end{flushleft}

% Add optional History. Make this a compact list
\ifshowhistory{
\setitemize{noitemsep,topsep=0pt,parsep=0pt,partopsep=0pt}  
\begin{flushleft}
\vspace{0.25cm}
Memo history:
\begin{itemize}
\itemsep0em
\memohistory{}
\end{itemize}
\end{flushleft}
}\fi

%%%%%%%%%%%%%%%%%%%%%%%%%%%%%%%%
%%%%%  Optional Abstract   %%%%% 
%%%%%%%%%%%%%%%%%%%%%%%%%%%%%%%%  
                                 
\begin{abstract} 

Starting from the observing campaign in April 2018, the Greenland Telescope (GLT) has been added as a new station of the Event Horizon Telescope (EHT) array. Visibilities on baselines to GLT, particularly in the North-South direction, potentially provide valuable new constraints for the modeling and imaging of sources such as M87*. The GLT's location at high Northern latitudes adds unique challenges to its calibration strategies. Additionally, the performance of the GLT was not optimal during the 2018 observations due to it being only partially commissioned at the time. This document describes the steps taken to estimate the various parameters (and their uncertainties) required for the absolute flux calibration of the GLT data as part of the EHT. In particular, we consider the non-optimized status of the GLT in 2018, as well as its improved performance during the 2021 EHT campaign. 

\end{abstract} 

\newpage
 
\tableofcontents

%%%%%%%%%%%%%%%%%%%%%%%%%%%%%
%%%%%  Start Report   %%%%% 
%%%%%%%%%%%%%%%%%%%%%%%%%%%%%

% Optional style settings 
% Add empty line in front of paragraphs, no indent, and global compact lists
\setlength{\parskip}{\baselineskip}
\parindent 0pt
\vspace{1.0cm}
%\newpage

\newcommand{\sm}[1]{\textcolor{blue}{SM: #1}}
\newcommand{\kk}[1]{\textcolor{red}{KK: #1}}

\section{Introduction}\label{introduction}

The Greenland Telescope \citep[GLT, ][]{inoueetal14,chenetal23} is a 12-m single-dish (sub-)mm telescope currently located at the Thule Air Base in Greenland (LAT: $+76^{\circ}32\arcmin 06 \farcs 6$, LON: $-68^{\circ}41\arcmin 08 \farcs 8$). It is managed and operated by the Academia Sinica Institute of Astronomy and Astrophysics (Taiwan) and the Smithsonian Astrophysical Observatory (USA). The antenna was originally the North American ALMA prototype antenna \citep{mangumetal06a,mangumetal06b}, and has since been retro-fitted to withstand the harsh Arctic conditions \citep{raffinetal14,raffinetal16}. While its current altitude is close to sea level (87.4\,m), the eventual goal is to move the telescope to the Summit station in Greenland at an altitude of 3200\,m, where even THz observations are possible \citep{matsushitaetal17}. Despite the low altitude at Thule, 230\,GHz observations can still be conducted due to the dry conditions at the site, with a mean atmospheric opacity of $\tau = 0.17$ at zenith during the winter season \citep{matsushitaetal22}.

The construction and re-assembly of the GLT was completed in July 2017 \citep{chenetal18,chenetal23}, after which the frontend and backend systems were installed \citep{kuboetal18,hanetal18}, together with the computer and network setup \citep{nishiokaetal18,huangetal20}. Scientific commissioning began in December 2017 \citep{matsushitaetal18}. The first ALMA-GLT fringes were detected for the quasar 3C279 as part of the EHT dress rehearsal in January 2018. The GLT was included in the EHT array for the April 2018 observing campaign, which represents the GLT's first scientific observations \citep{matsushitaetal18}, despite it being only partially commissioned. Commissioning activities continued through 2019 and 2020, during which the EHT observing campaigns were cancelled due to technical issues and the global Covid-19 pandemic. Transitioning to remote operations during the pandemic allowed the GLT to participate in the EHT observing campaign in April 2021.

\subsection{GLT-Specific Calibration Challenges}\label{challenges}

The GLT's location at high latitude and extreme weather conditions place unique challenges for GLT flux calibration. These are:

\begin{enumerate}
\item \textbf{Limited variations in elevation of available calibrator sources} due to the telescope's extreme Northern location means that it is difficult to obtain gain curves by observing a source over a large range in source elevation.

\item \textbf{Limited availability of bright sources with known brightness temperatures}, namely planets, due again to the telescope's high latitude location. As the Sun does not rise above the horizon in winter, the availability of planets, located along or close to the ecliptic, is very limited during winter.

\item \textbf{Significant differences between the atmospheric temperature and the ambient temperature at the receiver}, due to the receiver cabin being heated to room temperatures (typically set to $\sim 16^{\circ}$C) for ideal performance of the electronics. This may lead to large systematic errors in the estimation of the effective system temperatures of the telescope based on the standard voltage measurements of hot load against that of the cold sky. 
\end{enumerate}

\subsection{GLT Performance Issues During the 2018 EHT Observations}\label{status2018}

As mentioned, the performance of the GLT was not optimized during the 2018 observations. Remaining issues at the time included:
\begin{enumerate}
	\item \textbf{Low antenna efficiency.} Photogrammetry procedures conducted in the summer of 2018 found the dish to have a poor surface accuracy of $\sim 180\, \mu \rm m$ and a systematic saddle-shaped curvature (Figure~\ref{fig:surface}, left). This was most likely imprinted when the dish was lifted by crane to be mounted on the support cone, since the curvature matched the locations at which the harnesses were attached \citep{koayetal20, chenetal23}. 
	
	\begin{figure} [t]
 	\begin{center}
 		\begin{tabular}{c} %% tabular useful for creating an array of images 
 			\includegraphics[width=8.2cm]{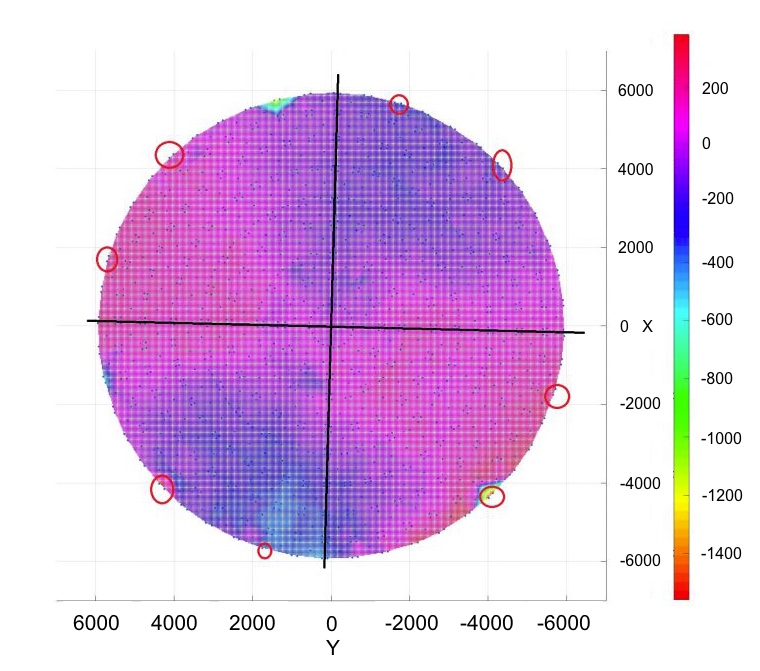}
 			\includegraphics[width=9.7cm]{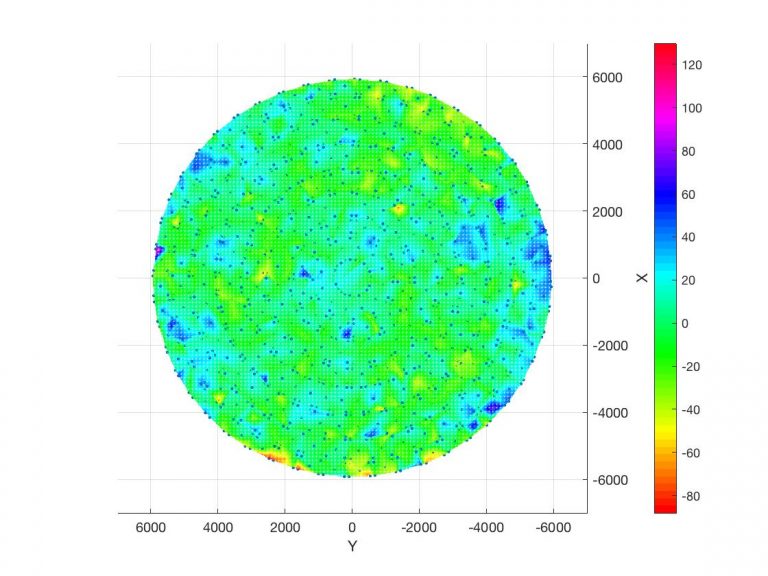}
 		\end{tabular}
 	\end{center}
 	%>>>> use \label inside caption to get Fig. number with \ref{}
 	\caption{
 	{\label{fig:surface} Left: Antenna rms surface accuracy of $\sim 180 \, \mu \rm m$ measured in summer 2018, likely to be the state of the dish during the EHT observations. The saddle-shape is consistent with the location of the attached harnesses (shown as red ellipses) when the dish was lifted onto the telescope mount. Right: Antenna rms surface accuracy of $\sim 25 \, \mu \rm m$ measured after panel adjustments conducted between August and October 2018 \citep{koayetal20}.}}
 \end{figure} 
	
	\item \textbf{Continuum detector instability and amplitude offset.} Measurements of the system noise temperature, antenna efficiency and the sky opacity (using sky-dipping) for the GLT during the April 2018 observations were made using the continuum detector, whose output was later found to exhibit small fluctuations (2.8\% peak-to-peak) resulting from its sensitivity to receiver cabin ambient temperatures. Concurrent effective system noise temperature and sky-dipping atmospheric opacity measurements using both the continuum detector and the power-meter, conducted in October 2018, reveal time-dependent amplitude offsets between both measurements due to the continuum output variability. A rescaling/recalibration of the April 2018 continuum detector output is needed for the estimation of the system noise temperatures.
	
	\item \textbf{Large pointing errors at low elevation.} The low sensitivity of the GLT at the time led to very low SNR observations of pointing sources at low elevations. The pointing model derived just before the EHT observations was found to have pointing errors of $10 - 30 \arcsec$ at source elevations $\lesssim 10^{\circ}$. This leads to additional amplitude losses for sources observed at these low elevations.
	
\end{enumerate}

Another challenge for the calibration of the 2018 GLT data is that continued commissioning activities were carried out through the summer of 2018 and beyond. The performance and characteristics of the GLT have thus changed significantly multiple times since the EHT 2018 observations. For example, the antenna efficiency, beam shape and gains have changed, and their values in April 2018 can only be estimated using archival measurements close to those dates. If an issue is found in the measurements, there is no way of `re-setting' the telescope conditions to those in April 2018 for further measurements and verification. 

\subsection{GLT Status During the 2021 EHT Observations}\label{status2021}

The issues pertaining to the 2018 GLT data have since been rectified:
\begin{enumerate}

\item \textbf{The antenna efficiency has significantly improved} from $\sim 22$\% in 2018 to $\sim 65$\% at 230\,GHz, via photogrammetry and subsequent panel adjustments. The surface accuracy of the GLT is now at $\sim 25 \, \mu \rm m$ in summer and $\sim 40 \, \mu \rm m$ in winter (Figure~\ref{fig:surface}, right; \citealt{koayetal20}). 

\item \textbf{The more stable outputs of the power-meter} (no more than 0.5\% peak-to-peak variations) are now used for measurements of the system temperatures and antenna efficiencies. 

\item \textbf{The pointing accuracy at low elevations has improved}. With the improved antenna sensitivity, we were able to track pointing sources down to very low elevations with high SNR detections, providing better constraints for the pointing model at these elevations. The pointing uncertainties are now typically at $\sim 3\arcsec$ at elevations as low as $7^{\circ}$ at 230\,GHz.

\end{enumerate}

On the other hand, due to the pandemic, regularly scheduled maintenance of the telescope frontends, backends and other systems could not be carried out prior to the EHT 2021 observing campaign. Furthermore, the GLT observations as part of the EHT 2021 campaign were conducted via remote operations for the very first time. Some issues affecting the flux calibration of the GLT data during the 2021 campaign include the following:

\begin{enumerate}

\item Due to a technical problem with the block down-converter, the $T_{\rm sys}^{*}$ measurements obtained on one of the observing days, 13 April, were erroneous and cannot be used. This problem was fixed and $T_{\rm sys}^{*}$ could be measured properly for the rest of the campaign.

\item The on-site 225 GHz tipping radiometer broke down in early March 2021 due to hardware problems, so there was no opacity monitoring during the EHT 2021 observations. To make up for this, we performed sky dipping measurements with the GLT to determine the sky opacity before and after each observing track, and whenever possible, in between scans separated by large gaps in time. 

\item Due to poor weather conditions leading up to the observations, the antenna efficiency at 345\,GHz was not measured in 2021.
\end{enumerate}

\subsection{Scope of this Memo}\label{scope}

All the issues outlined above for the GLT in general and those pertaining specifically to the 2018 and 2021 EHT observations require that additional steps and care be taken when preparing the meta-data for the absolute flux calibration of the GLT data. This document describes these steps in detail. In the next section (Section~\ref{concepts}), we provide a brief overview of the standard telescope parameters necessary for the absolute flux calibration of (sub-)mm VLBI data. In Section~\ref{tsys}, we describe the measurement and derivations of the time-dependent effective system temperatures for the GLT, including the specific fixes needed for the 2018 measurements due to the continuum detector instability. We then discuss the estimation of the antenna efficiencies in Section~\ref{AeffDPFU} before and after the antenna surface adjustments; this is followed by a brief remark on the antenna gains (Section~\ref{gaincurve}). Section~\ref{pointing} discusses the estimated pointing uncertainties and their effects on the flux calibration. The performance of the GLT during the EHT 2018 and 2021 campaigns is summarized in Section~\ref{summary}.

\section{Basic Concepts for (Sub-)mm VLBI and EHT Flux Calibration}\label{concepts}

Detailed elucidations of the basic terms and concepts for a-priori flux calibration of (sub-)mm VLBI and EHT data are provided in the EHT memos by \citet{issaounetal17} and \citet{janssenetal19}. In this section, we provide a brief summary of these terms, mainly those that are referred to in this memo.

To convert VLBI visibility amplitudes from correlation coefficients to units of flux density, taking into account the inhomogeneity of telescopes in the array, the system equivalent flux densities (SEFD) of each telescope need to be determined. For a single dish telescope, the SEFD is given by \citep[e.g.,][]{ehtpaperIII19}:
\begin{equation}\label{eqSEFD}
{\rm SEFD} = \dfrac{T_{\rm sys}^{*}}{{\rm DPFU} \times g_E}~,
\end{equation}
where the degrees per flux density unit (DPFU) is a conversion factor from units of temperature (K) to units of flux density (Jy), and $g_E$ is the elevation-dependent gain curve of the telescope. The DPFU can be estimated as:
\begin{equation}\label{eqDPFU}
{\rm DPFU} = \dfrac{\eta_{\rm A} A_{\rm geom}}{2k}~,
\end{equation}
where $\eta_{\rm A}$ is the aperture efficiency, $A_{\rm geom}$ is the geometrical area of the antenna dish, and $k = 1.38 \times 10^{3}$ is the Boltzmann constant in units of Jy\,m$^2$/K.

$T_{\rm sys}^{*}$ is the effective system noise temperature \citep{issaounetal17,janssenetal19}, corrected for atmospheric attenuation. This needs to be differentiated from the regular $T_{\rm sys}$ value used in cm-wavelength VLBI which does not account for the atmospheric attenuation. $T_{\rm sys}^{*}$ is thus given by:
\begin{equation}\label{eqTsys}
T_{\rm sys}^{*} = \frac{e^{\tau}}{\eta_{\rm l}}T_{\rm sys}~, 
\end{equation}
where $\tau$ is the atmospheric opacity, and $\eta_{\rm l}$ is the forward efficiency (i.e., the coupling of the receiver to the sky, accounting for ohmic losses and rearward spillover and scattering losses). Single dish (sub-)mm telescopes typically measure $T_{\rm sys}^{*}$ directly by placing a hot load of known temperature in the signal path. This is done regularly during or in between target scans to account for the time and elevation (airmass) dependence of $T_{\rm sys}^{*}$ as the telescope points to different targets.

The above parameters, i.e., $T_{\rm sys}^{*}$, DPFU and $\eta_{\rm el}$, need to be measured for the GLT to enable the absolute flux calibration of the visibility data on GLT baselines. These parameters are typically stored in a standard format calibration table (ANTAB), which is then used by data reduction software to calculate the SEFDs and apply the calibrations.

\section{Estimating Effective System Temperature}\label{tsys}

 To calibrate the visibility amplitude for the EHT data, it is necessary to measure the $T_{\rm sys}^{*}$ as a function of time/elevation for each station in the array during the observing track. It is typically assumed that the ambient, the atmospheric, and the hot load temperatures are similar enough such that the process to derive $T_{\rm sys}^{*}$ values can be simplified. However, such an assumption may lead to significant systematic errors in the estimates of the GLT $T_{\rm sys}^{*}$, for which the effective atmospheric temperature $T_{\rm atm}$ is much lower than the ambient temperature $T_{\rm amb}$ at the receiver and hot load, compared to many of the other EHT sites. This is because the receiver cabin is heated to room temperatures for the proper functioning of the electronics systems. Therefore, a more generic equation which allows $T_{\rm amb}$ and $T_{\rm atm}$ to be different is needed to estimate $T_{\rm sys}^{*}$ for the GLT and reduce the systematic uncertainties. In this section, we present the more generic formulation of $T_{\rm sys}^{*}$, and provide an estimate of the systematic error in $T_{\rm sys}^{*}$ that arises if we ignore the differences between $T_{\rm amb}$ and $T_{\rm atm}$ for the GLT.

\subsection{A More Generic Formulation for $T_{\rm sys}^{*}$}\label{tsysest}

The system noise temperature, $T_{\rm sys}$, describes the noise contribution from the receiver and the sky to the flux density measurement of a radio source. For the case where the effective atmospheric temperature ($T_{\rm atm}$) is comparable to the ambient (or room) temperature of the receiver cabin ($T_{\rm amb}$), $T_{\rm sys}$ is expressed following the nomenclature used in \citet{issaounetal17} as:
\begin{equation}\label{eqtsys}
T_{\rm sys} = T_{\rm rx} + T_{\rm sky} = T_{\rm rx} + T_{\rm atm}(1 - \eta_{\rm l}e^{-\tau})~,
\end{equation}
 where $T_{\rm rx}$ refers to the receiver noise temperature, and $T_{\rm sky} = T_{\rm atm}(1 - e^{-\tau})$ is the sky brightness temperature. Again, $\eta_{\rm l}$ is the forward efficiency and $\tau$ is the line-of-sight atmospheric opacity.  If $T_{\rm atm}$ signficantly differs from $T_{\rm amb}$, for a station such as the GLT, then the system temperature is expressed more generally as \citep{ulichandhaas76, rohlfsandwilson06}:

\begin{align}\label{eqtsysgen}
T_{\rm sys}  & = T_{\rm rx} + \eta_{\rm l}T_{\rm sky} + (1-\eta_{\rm l})T_{\rm amb} \nonumber\\
    & = T_{\rm rx} + \eta_{\rm l}T_{\rm atm}(1 - \eta_{\rm l}e^{-\tau}) + (1-\eta_{\rm l})T_{\rm amb} \,.
\end{align}
%assuming $\eta_{\rm l} = 1$.
%\textcolor{teal}{Are we then saying that if $T_{\rm atm} = T_{\rm amb}$ then $T_{\rm sky} = T_{\rm atm}(1 - \eta_{\rm l}e^{-\tau})$, but if $T_{\rm atm} \neq T_{\rm amb}$ then $T_{\rm sky} = T_{\rm atm}(1 - e^{-\tau})$?, Shouldn't $T_{\rm sky} = T_{\rm atm}\eta_{\rm l}(1-{\rm e}^{-\tau}) + T_{\rm amb}(1-\eta_{\rm l})$ instead when $T_{\rm atm} \neq T_{\rm amb}$? Otherwise $T_{\rm sys} \neq T_{\rm rx} + T_{\rm sky}$ in general. If this is not intended, then the first definition needs to be changed to $T_{\rm sys} \neq T_{\rm rx} + T_{\rm sky} + T_{\rm spillover}$, or?}

To further account for the signal attenuation caused by the atmosphere and the rearward efficiency loss, it is convenient to define the effective system noise temperature, $T_{\rm sys}^{*}$ (Section~\ref{concepts}), which explicitly indicates that the system sensitivity drops exponentially as line-of-sight opacity increases:
\begin{equation}
T_{\rm sys}^{*} = \frac{e^{\tau}}{\eta_{\rm l}} T_{\rm sys} = \frac{e^{\tau}}{\eta_{\rm l}}[ T_{\rm rx} + \eta_{\rm l}T_{\rm sky} + (1-\eta_{\rm l})T_{\rm amb} ]~.
\end{equation}

To determine $T_{\rm sys}^{*}$, the receiver output voltage $C_{\rm amb}$ is first measured when the hot load with an ambient temperature, $T_{\rm amb}$, is placed in the signal path of the receiver, and this load blocks everything except for the receiver noise. This step is followed by the measurement of the receiver voltage $C_{\rm sky}$ when the hot load is removed and the feed horn is directly pointing towards the cold sky with a temperature of $T_{\rm sky}$ at a certain elevation. For the hot load measurement, one can relate the receiver output and the ambient/receiver temperature as:
\begin{equation}
C_{\rm amb} = G(T_{\rm amb} + T_{\rm rx})~,
\end{equation}
 where $G$ is the voltage gain. For the cold sky measurement, the receiver responds to empty sky noise ($T_{\rm sky}$) and noise from the receiver cabin (corresponding to $T_{\rm amb}$). In this case, the relationship between the receiver output and the sky/ground noise can be expressed as (see Equation 7.26 in \citet{rohlfsandwilson06}):
\begin{equation}
C_{\rm sky} = G\left[\eta_{\rm l}T_{\rm sky} + (1-\eta_{\rm l})T_{\rm amb} + T_{\rm rx}\right]~.
\end{equation}
Taking the difference between $C_{\rm amb}$ and $C_{\rm sky}$, we obtain:
\begin{equation}
\Delta C_{\rm cal} \equiv C_{\rm amb} - C_{\rm sky} =  G\eta_{\rm l}(T_{\rm amb} - T_{\rm sky})~.
\end{equation}
Since $T_{\rm sky} = T_{\rm atm}(1 - e^{-\tau})$, $\Delta C_{\rm cal}$ can be expressed as:
\begin{equation}\label{eqdeltac}
\Delta C_{\rm cal} = G\eta_{\rm l}(T_{\rm amb} - T_{\rm atm} + T_{\rm atm} e^{-\tau})~.
\end{equation}

For many of the EHT stations, it is reasonable to assume that $T_{\rm amb} = T_{\rm atm}$, such that $\Delta C_{\rm cal}$ $=$ $G\eta_{\rm l}T_{\rm amb}e^{-\tau}$. As mentioned, the difference between $T_{\rm amb}$ and $T_{\rm atm}$ is non-negligible for the GLT; in this case, one can express the ambient temperature as $T_{\rm amb}$ $=$ $T_{\rm atm}$ $+$ $\Delta T$, where  $\Delta T$ is the difference between the ambient and the effective atmospheric temperature, and Equation~\ref{eqdeltac} can be re-written as:
\begin{equation}
\Delta C_{\rm cal} = G\eta_{\rm l}(\Delta T + T_{\rm atm} e^{-\tau})~.
\end{equation}

Now, in order to relate $T_{\rm sys}^{*}$ to the measured receiver outputs (i.e., $C_{\rm sky}$ and $C_{\rm amb}$), we evaluate the ratio, $r_{\rm cal}$, of $C_{\rm sky}$ to $\Delta C_{\rm cal}$, obtaining:
\begin{equation}\label{eqrcal}
r_{\rm cal} \equiv { C_{\rm sky} \over \Delta C_{\rm cal}} = { G[\eta_{\rm l}T_{\rm sky} + (1-\eta_{\rm l})T_{\rm amb} + T_{\rm rx}]  \over G\eta_{\rm l}(\Delta T + T_{\rm atm} e^{-\tau})}~,
%=  { \eta_{\rm l}T_{\rm sky} + (1-\eta_{\rm l})T_{\rm cabin} + T_{\rm rx}  \over \eta_{\rm l}(\Delta T + T_{\rm atm} e^{-\tau})}~.
\end{equation}
which can be rearranged as:
\begin{align}\label{eqrcal2}
r_{\rm cal} & = { \eta_{\rm l}T_{\rm sky} + (1-\eta_{\rm l})T_{\rm amb} + T_{\rm rx}  \over \eta_{\rm l}T_{\rm atm} e^{-\tau}[1+\Delta T/(T_{\rm atm} e^{-\tau}) ] } \nonumber \\
                 & =  { T_{\rm sys}^{*} \over T_{\rm atm}[1+(\Delta T/T_{\rm atm}) e^{\tau} ] }~.
\end{align}
Finally, one obtains:
\begin{equation}\label{eqtsysglt}
T_{\rm sys}^{*} = r_{\rm cal} T_{\rm atm}[1+(\Delta T/T_{\rm atm}) e^{\tau}] ~.
\end{equation}

In summary, the determination of the effective system temperature of a submillimeter radio telescope involves the measurements of the voltage outputs from the receiver which can be used to derive $T_{\rm sys}^{*}$ via equation~\ref{eqtsysglt}. This equation explicitly accounts for the difference between $T_{\rm atm}$ and $T_{\rm amb}$, and thus should be applied for GLT $T_{\rm sys}^{*}$ measurements.

\subsection{$T_{\rm sys}^{*}$ Errors When Not Using Generic Equation}

For telescopes at which $\Delta T = 0$ is a reasonable assumption, the ratio $r_{\rm cal}$ is simply the effective system temperature $T_{\rm sys}^{*}$ divided by $T_{\rm amb}$, since from Equation~\ref{eqrcal}:
\begin{equation}
r_{\rm cal, \Delta T = 0} =  \frac{e^{\tau}}{\eta_{\rm l}} \left( {\eta_{\rm l}T_{\rm sky} + (1-\eta_{\rm l})T_{\rm amb} + T_{\rm rx} \over T_{\rm atm} } \right) = {T_{\rm sys}^{*} \over T_{\rm amb} }~.
\end{equation}
In this case, one can easily evaluate $T_{\rm sys}^{*}$. To emphasize that this expression is valid only when $\Delta T = 0$, we define the effective system temperature when $\Delta T = 0$ as $T_{\rm sys0}^{*}$: 
\begin{equation}
T_{\rm sys0}^{*} \equiv r_{\rm cal, \Delta T = 0}T_{\rm amb}~.
\end{equation}
 
Substituting $T_{\rm atm} = T_{\rm amb} - \Delta T$ into Equation~\ref{eqtsysglt}, we have:
\begin{equation}
T_{\rm sys}^{*} = r_{\rm cal} (T_{\rm amb}-\Delta T)(1+\Delta T  e^{\tau}/(T_{\rm amb}-\Delta T)) ~,                        
\end{equation}
which, to first order, can be approximated as:
\begin{align}
T_{\rm sys}^{*} & \approx r_{\rm cal}T_{\rm amb}\left[1+\frac{\Delta T}{T_{\rm amb}}(e^{\tau_{\nu}}-1)\right] \nonumber \\
& =  T_{\rm sys0}^{*}\left[1+\frac{\Delta T}{T_{\rm amb}}(e^{\tau_{\nu}}-1)\right] ~.
\end{align}
The terms in the square parenthesis in the above equation represent the correction factor that needs to be multiplied to the $T_{\rm sys}^{*}$ values evaluated assuming $T_{\rm amb}$ $=$ $T_{\rm atm}$. Considering typical receiver cabin temperatures of $15^{\circ}$C, and typical winter atmospheric temperatures of $-25$ to $-35^{\circ}$C, $\Delta T$ can be as large as 50$^{\circ}$K.  Assuming $T_{\rm amb}$ $=$ 290$^{\circ}$K and $\tau$ $=$ 0.2, then this correction factor represents a 4\% error relative to $T_{\rm sys0}^{*}$, if only $T_{\rm amb}$ is used to calculate $T_{\rm sys}^{*}$.  If $T_{\rm atm}$ is used instead to calculate $T_{\rm sys0}^{*}$, i.e. if one had assumed $T_{\rm sys0}^{*} \equiv r_{\rm cal, \Delta T = 0}T_{\rm atm}$, the error in $T_{\rm sys}^{*}$ would be as large as 25\%, based on the term in parenthesis in Equation~\ref{eqtsysglt}. 

Therefore, for typical conditions at the GLT, $T_{\rm amb}$ should used instead of $T_{\rm atm}$ in the calculation of $T_{\rm sys}^{*}$ , when using the standard equation. In such a case, the $T_{\rm sys}^{*}$ systematic uncertainties introduced would be no more than 4\%. However, to reduce these systematic uncertainties further, we use Equation~\ref{eqtsysglt} to estimate the GLT $T_{\rm sys}^{*}$ values.

\subsection{System Noise and Temperature Measurements at the GLT}\label{tsysmeas}

In order to measure $T_{\rm sys}^{*}$, a continuum detector was installed on the GLT. A 4\,GHz bandwidth intermediate frequency (IF) signal (at 3.85 - 7.85 GHz) is divided with a coupler in the GLT IF assembly system, and the continuum detector measures the power of this signal. It was later discovered that the continuum detector output was sensitive to the receiver cabin ambient temperature, which was unstable during the 2018 EHT campaign due to insufficient optimization of the HVAC receiver cabin temperature control (see Sect.~\ref{contrescale}). After the 2018 EHT observations, we have permanently switched to using the power-meter for the $T_{\rm sys}^{*}$ measurements, since it provides more stable outputs that are less dependent on the ambient temperature.

For the EHT observations, the GLT measures the system temperatures before and after each scan, using the continuum detector in 2018, and the power-meter from 2021. During a $T_{\rm sys}^{*}$ measurement, the system noise of the hot load ($C_{\rm amb}$) is first measured by the continuum detector for a duration of 10\,s, followed by that of the sky ($C_{\rm sky}$) for 10\,s. For the 2018 observations, the receiver cabin temperature ($T_{\rm amb}$) is automatically recorded by 5 separate sensors located at different positions in the cabin, and their mean values are assumed to be equivalent to the hot load temperature. A new sensor has since been installed on the hot load, so for the 2021 observations onwards, $T_{\rm amb}$ is directly read from this hot load temperature sensor. Comparing the hot load temperature readings with the mean temperatures from the five cabin sensors show differences of no more than a few degrees K ($< 1\%$). Therefore the 2018 $T_{\rm amb}$ measurements are expected to be representative of the hot load temperatures.    

The atmospheric opacity is measured with the GLT using the sky-dipping technique, as well as with an on-site 225 GHz tipping radiometer \citep[]{matsushitaetal17, matsushitaetal22}, which as noted, was not functioning during the 2021 EHT observing campaign. The effective atmospheric temperature, $T_{\rm atm}$, is also monitored continuously by an on-site weather station.

The method described in this section was used to determine the GLT $T_{\rm sys}^{*}$ for each scan during the 2021 observations, and will be used for all subsequent EHT observations. For the 2018 EHT observations, $C_{\rm amb}$ and $C_{\rm sky}$ were also measured using the method described above, but the $T_{\rm sys}^{*}$ estimates derived are unreliable due to the 2.8\% fluctuations of the continuum detector as a function of time and cabin temperature. We used a workaround method to derive the $T_{\rm sys}^{*}$ values for 2018, which we describe in the next subsection.

\subsection{Re-scaling the Continuum Detector Outputs for the 2018 Observations}\label{contrescale}
	
\begin{figure} [t]
 	\begin{center}
 		\begin{tabular}{c} %% tabular useful for creating an array of images 
 			\includegraphics[width=12cm]{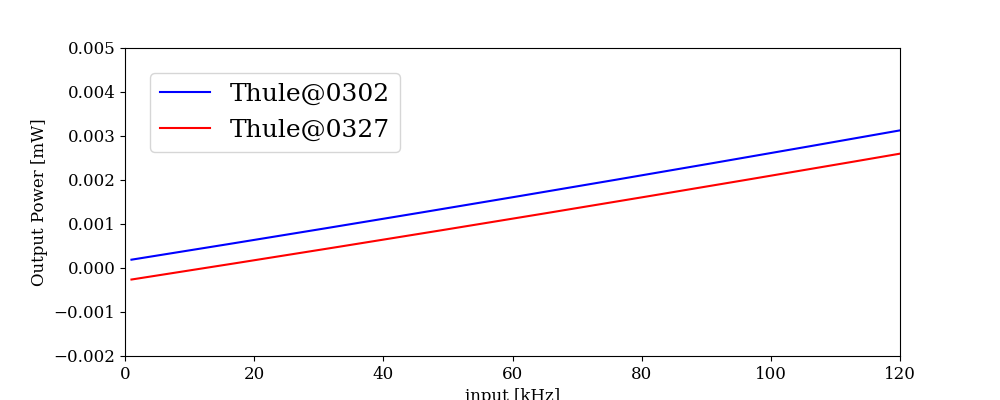}
 		\end{tabular}
 	\end{center}
 	%>>>> use \label inside caption to get Fig. number with \ref{}
 	\caption{
 	{\label{fig:CDcurve} Estimated calibration curves for the GLT continuum detector; these were measured by comparing the input power level of the noise source measured with a power-meter and output of the continuum detector, as a function of frequency. While calibration curves are expected to have first and second order terms, the frequency range we were using was mainly dominated by the zeroth order terms. Therefore, the relationship between the power-meter and continuum detector measurements is expected to vary linearly in frequency, and to require only a constant offset in the observed output power. }}
 \end{figure}

We discovered that the continuum detector output showed 2.8\% fluctuations that were consistent with the receiver cabin temperature variations linked to the HVAC cycles. The continuum detector output is read out as a function of frequency. This output is then converted into units of power using the apriori obtained calibration curve. Figure \ref{fig:CDcurve} shows the calibration curves obtained at two different days before the EHT observations in 2018, which are different for each measurement, presumably because of the temperature fluctuations. We thus expect the continuum detector measurements to have large uncertainties.

Therefore, to estimate the GLT $T_{\rm sys}^{*}$ values for 2018, we (1) estimate the nominal receiver temperature, then (2) estimate  the atmospheric temperatures and zenith opacities measured by the 225 GHz tipping radiometer, and (3) estimate the effective system temperature by considering the elevation towards the target source. We now describe each of these steps in detail.

\subsubsection{Estimation of the nominal receiver temperature}

\begin{figure} [t]
        \begin{center}
 		\begin{tabular}{c} %% tabular useful for creating an array of images 
 			\includegraphics[width=12cm]{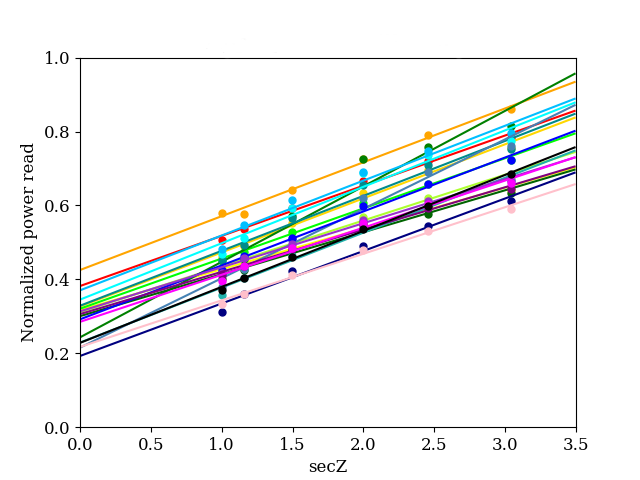}
 		\end{tabular}
 	\end{center}
 	%>>>> use \label inside caption to get Fig. number with \ref{}
 	\caption{
 	{\label{fig:secZs} Tipping measurements for the GLT during the 2018 EHT campaign, with each colored line representing one of the 20 recorded measurements.}}
 \end{figure} 

We estimate the receiver temperature ($T_{\rm rx}$) using sky dipping observations. The output power from the continuum detector can be expressed as:
\begin{equation}
P_{\rm out} = G\left(T_{\rm rx} + T_{\rm sky}\right) = G\left(T_{\rm rx} + T_{\rm atm}\left(1 - e^{-\tau}\right)\right)~.
\end{equation}
Assuming a plane-parallel atmosphere, the opacity, $\tau$, can be expressed as $\tau_{0}\,{\rm sec}(Z)$, where $\tau_{0}$ is the opacity towards the zenith and $Z$ is the elevation angle, giving:  
\begin{equation}\label{eqsecZ}
P_{\rm out} = G\left(T_{\rm rx} + T_{\rm atm}\left(1 - e^{-\tau_{0}\,{\rm sec}(Z)}\right)\right)~.
\end{equation}
Normalizing Equation~\ref{eqsecZ} by the output measured towards the hot load, we obtain: 
\begin{equation}\label{eqsecZnorm}
P_{\rm nor}= \frac{\left(T_{\rm rx} + T_{\rm atm}\left(1 - e^{-\tau_{0} {\rm sec} (Z)}\right)\right)}{T_{\rm rx} + T_{\rm amb}}~.
\end{equation}

\begin{figure} [t]
 	\begin{center}
 		\begin{tabular}{c} %% tabular useful for creating an array of images 
 			\includegraphics[width=12cm]{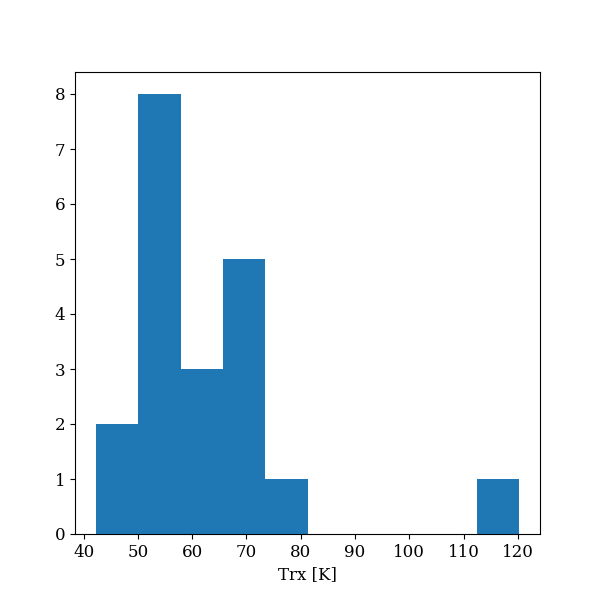}
 		\end{tabular}
 	\end{center}
 	%>>>> use \label inside caption to get Fig. number with \ref{}
 	\caption{
 	{\label{fig:Trx} Histogram of the estimated receiver temperature based on sky dipping observations.  }}
 \end{figure}

There were twenty sky dipping observations at 230\,GHz after the continuum detector was last calibrated on 2018 March 27.  In Figure~\ref{fig:secZs}, we show all of the sky dipping measurements. The horizontal axis corresponds to the path length (air mass) and the vertical axis corresponds to the output power normalized by the output power measured towards the hot load, $P_{\rm nor}$. In the tipping curve, the slope corresponds to the increase of the sky temperature at different air masses, which should be different between measurements due to the varying opacity and atmospheric temperature. The receiver temperature is expected to correspond to the normalized power at the cross section with the vertical axis, when ${\rm sec}(Z) = 0$, such that:
\begin{equation}\label{eqsecZ0norm}
P_{\rm nor, sec(Z) = 0} = \frac{T_{\rm rx}}{T_{\rm rx} + T_{\rm amb}}~.
\end{equation}
We expect this cross-section to be a constant value that is dependent only on $T_{\rm rx}$ and $T_{\rm amb}$, both of which should not vary significantly. However, we clearly see that the cross sections also show large fluctuations presumably due to the fluctuations of the output of the continuum detector. In Figure~\ref{fig:Trx}, we show the histogram of the estimated receiver temperature based on each sky dipping observation. These receiver temperatures are estimated from the measured normalized output power using Equation~\ref{eqsecZ0norm}. Most of the measurements are clustered in the range between 40 to 80 K.  While individual estimated numbers are not reliable due to the continuum detector output fluctuations, we used the median of those measurements as the nominal value for the GLT receiver temperature in the 2018 experiments, which is $57.9 \pm 9.2$\,K. The estimated uncertainty excludes the outlier value at about 115\,K.

\subsubsection{Estimation of the sky opacity}

During the EHT 2018 observations, tipping measurements are conducted every 10 min with the 225 GHz radiometer. For each tipping, we measure the brightness temperature of the atmosphere at 5 different elevation angles, tilting in both the Northern and Southern directions. Elevation angles are 90, 42, 30, 24.6 and 19.2 degrees, corresponding to zenith air masses of 1, 1.5, 2, 2.5 and 3, respectively. We note that the opacity data will be used when both the opacities toward Northern and Southern directions are the same within errors. We also measure a built-in absorber brightness and monitor the atmospheric temperature in order to do a gain calibration at the time of tipping. By utilizing those measurements, we estimated the opacity towards the zenith $\tau_{0}$ every 10 min (Matsushita et al., in prep.).
%\textcolor{green}{Note that our radiometer was tipping towards the Southern and Northern directions, while the actual observations are pointed towards the Southeast through Southwest directions throughout the observing track. These differences in the sampled direction could lead to additional uncertainties in the derivation of $\tau$.} \textcolor{red}{Here, we only accept the opacities when both the Northern and Southern directions are the same, so the green text part can be deleted.} 
%\textcolor{red}{(This explanation is not correct. The radiometer output in the gltmonitor is measured every second with the mirror pointing at zenith, and estimate the opacity using the gain value that measured every half year by Pierre with the liquid nitrogen calibration)}
%<== Turned out that we used the properly analyzed opacity values mentioned above, not the real-time measurement values. So, the explanation is correct.

\subsubsection{Estimation of the effective system temperature}

Based on the estimates of $\tau_{0}$ and $T_{\rm rx}$, and measurements of $T_{\rm atm}$ and $T_{\rm atm}$, we calculated the effective system temperature for each scan. Since the opacities and temperatures were measured every 10 min and there are missing data in some cases, we just used the values closest in time to each particular scan. We obtained $\tau$ values towards the source by considering the elevation of the telescope during that scan, which was extracted from the observing schedule file (VEX file) used to control the telescope. Based on the $T_{\rm rx}$ uncertainties and using nominal values for $\tau_{0}$ and $T_{\rm atm}$, we estimate $T_{\rm sys}^{*}$ uncertainties of $\sim 15\%$ for the GLT in 2018. This assumes that the continuum detector instabilities are the dominant sources of error in the $T_{\rm sys}^{*}$ estimate.
%We estimated sky temperature at starting and ending time for the each scan.

\section{Aperture Efficiency and DPFU}\label{AeffDPFU}

%how antenna efficiency is measured ... what equations are used
%could the differences in beam size in RCP and LCP be affected by the continuum detector instability differences between RCP and LCP? One of them is worse?

%%% 1. Forward spillover efficiency (Mangum 1993, PASP, 105, 117)
%%% 2. Aperture efficiency (Kraus 1986)
%%% 3. Planet temperature (Butler 2012, ALMA Memo 594)

By observing a source with known temperature and size, namely a planet, it is possible to derive the main-beam efficiency as follows \citep{mangum93}:
\begin{equation}
\eta_{\rm mb} = \frac{\frac{1}{2}T_{\rm A, planet}^{*}}{J(T_{\rm planet})-J(T_{\rm CMB})} \times \left\{1-\exp\left[-\ln2\left(\frac{\theta_{\rm planet}}{\theta_{\rm mb}}\right)^{2}\right]\right\}^{-1},
\label{eq_eta_mb}
\end{equation}
where:
\begin{itemize}
\item $T_{\rm A, planet}^{*}$ is the single-sideband antenna temperature of a planet
\item $J(T)$ is the Planck function at frequency $\nu$ and brightness temperature, given by: namely $J(T_{\rm B}) = \frac{h\nu/k}{\exp(h\nu/kT_{\rm B})-1}$
\item $T_{\rm planet}$ is the brightness temperatures of the planet
\item $T_{\rm CMB}$ is the brightness temperature of the cosmic microwave background
\item $\theta_{\rm planet}$ is the diameter of the planet in arcsec
\item $\theta_{\rm mb}$ is the telescope main-beam Full Width at Half Maximum (FWHM) in arcsec
\end{itemize}
In the typical case where the size of the observed planet is smaller or comparable to the FWHM of the telescope main-beam, this equation gives the main-beam efficiency (in the case where a planet is smaller than the main-beam, one needs to consider the filling factor within the main-beam). However, when the Moon instead of a planet is used for this measurement, since the size of the Moon is significantly larger than the main-beam and covers most of the side-lobes, the expression above actually gives the forward spillover efficiency, $\eta_{\rm fss}$ \citep{mangum93}. 

%\textcolor{red}{KK: From my understanding, they are slightly different. $\eta_l$ is the ratio of the power received in the forward direction (over 2$\pi$ steradian) relative to the power received in all directions (4$\pi$ steradian), it is the ohmic efficiency multiplied by the rearward spillover and scattering efficiency.}
%\textcolor{green}{SM: I think it is the same in reality, since we cannot measure 2$\pi$ steradian efficiency, but it is possible for Moon, and the sidelobe levels are so small outside Moon, so compared to other error budget, we can say the same.}

The main-beam efficiency can also be expressed as follows:
\begin{equation}
\eta_{\rm mb} = \frac{\theta_{\rm mb}^{2}A_{\rm eff}}{\lambda^{2}},
\end{equation}
where $A_{\rm eff}$ is the effective aperture area of the antenna. From this equation, $A_{\rm eff}$ can be expressed as:
\begin{equation}
A_{\rm eff} = \frac{\lambda^{2}\eta_{\rm mb}}{\theta_{\rm mb}^{2}}.
\label{eq_A_eff}
\end{equation}
The aperture efficiency, $\eta_{\rm A}$ is expressed as
\begin{equation}
\eta_{\rm A} = \frac{A_{\rm eff}}{A_{\rm geom}},
\end{equation}
where $A_{\rm geom}$ is the geometrical area of the antenna dish.
%physical aperture size.
Using Eqs.(\ref{eq_eta_mb}) and (\ref{eq_A_eff}), $\eta_{\rm A}$ can therefore be expressed as:
\begin{equation}
\eta_{\rm A} = \frac{\lambda^{2}T_{\rm A, planet}^{*}}{2\theta_{\rm mb}^{2}A_{\rm geom}[J(T_{\rm planet})-J(T_{\rm CMB})]} \times \left\{1-\exp\left[-\ln2\left(\frac{\theta_{\rm planet}}{\theta_{\rm mb}}\right)^{2}\right]\right\}^{-1}.
\end{equation}

We used the latest flux density models for planets in the ALMA Common Astronomy Software Application \citep[CASA;][]{mcmullinetal07} package \citep{butler12}. These are found in CASA versions 4.0 or later. The brightness temperatures of planets for the frequency range between 30 GHz and 1000 GHz are in the data files located at [CASA home directory]/data/alma/SolarSystemModels/. So far, we have only used Venus and Mars as the target planets for the GLT aperture efficiency measurements. The brightness temperature of Venus depends only on frequency, but that of Mars also depends on time.

In the case of Moon observations, the brightness temperature depends on the Moon's phase. The brightness temperature of the Moon, $T_{\rm B,\rm Moon}$, toward the center of the lunar disk is described as follows:
\begin{equation}
T_{\rm B,\rm Moon}(\lambda) = T_{0}(\lambda) + \frac{0.77T_{0}(\lambda)}{\sqrt{1 + 0.48\lambda + 0.11\lambda^{2}}} \cos[\chi - \xi(\lambda)],
\end{equation}
where
\[\chi \equiv \left(\frac{\rm Days\ since\ new\ Moon}{\rm Period\ of\ Moon}\right) \times 360 - 180,\]
\[\xi(\lambda) \equiv \tan^{-1}\left(\frac{0.24\lambda}{1 + 0.24\lambda}\right),\]
$\lambda$ is the observing wavelength, and the period of the Moon is 29.530589 days \citep{mangum93,krotikov64,linsky66,linsky73}.
Values for $T_{0}(\lambda)$ are $227.7\pm8.9$ K at 3 mm (100 GHz), $219.1\pm6.3$ K at 1 mm (300 GHz), and $217.6\pm8.1$ K at 300 $\mu$m (1000 GHz) \citep{linsky73}.

\subsection{GLT Aperture Efficiency and DPFU in 2018}\label{Aeff2018}

The $\eta_{\rm fss}$ at 230 GHz was observed around 1 AM on Mar.30, 2018 (UT), when it was 12.62 days after the new Moon. The results were 92.1\% for the Right-Hand Circular (RHC) polarization, and 92.2\% for the Left-Hand Circular (LHC) polarization.

The $\eta_{\rm A}$ for 230 GHz was observed on Mar.17, 2018 toward Venus at an elevation of 17$^{\circ}$, where there are no large pointing errors (see Section~\ref{pointing2018}). Due to the instability in the RCP data output, with large gain jumps during the scans of Venus, we could only use the LCP data. We obtained three separate measurements, and the results gave $\eta_{\rm A}$ of $21.6 \pm0.9\%$ ($\pm 1.4\%$ after adding the 5\% uncertainties of the Venus flux density model \citep{butler12}), which was much lower than the expected value for a dish surface accuracy of $40 \, \mu$m rms after panel adjustments but before the dish was transported to the site and mounted on the support cone. We later found that the low aperture efficiency was due to the surface deformation (saddle-like structure with the surface accuracy of $180 \, \mu$m, as shown in Figure~\ref{fig:surface}) when the dish was lifted using a crane \citep{koayetal20}. This led to significant astigmatism of the GLT antenna. These values of the antenna efficiency are thus zeroth order estimates, and their uncertainties are likely much larger due to the systematics introduced by the dish astigmatism. Additionally, the efficiency is derived based only on azimuthal scans of Venus. Altitudinal scans could not be performed due to the low elevation of the source, since there would be large variations in airmass throughout the scan.

Based on the aperture efficiency value of $21.6 \pm0.9\%$ at 230\,GHz, we estimate the DPFU using Equation~\ref{eqDPFU} to be $0.0088 \pm 0.0004$ K/Jy. Considering the quoted 5\%
uncertainties of the Venus flux density model \citep{butler12}, and adding that in quadrature to the measurement uncertainties gives us $0.0088 \pm 0.0006$ K/Jy. We use this DPFU value and uncertainty for both right and left hand circular polarizations.

\subsection{GLT Aperture Efficiency and DPFU in 2021}\label{Aeff2021}

%measurement details, antenna efficiency and DPFU for 2021
The $\eta_{\rm A}$ for 230 GHz was measured on March 6, 2020 toward Venus at elevations of $26^{\circ}-27^{\circ}$, just before the pandemic started world-wide. Both RCP and LCP receivers were working nominally at the time, and we took 13 individual measurements. The calculated aperture efficiencies are $67.1 \pm1.3\%$ and $64.7 \pm2.0\%$ for RCP and LCP, respectively. These values are consistent with the antenna surface accuracy of $40 \, \mu$m rms, assuming the geometrical antenna efficiency of $75\%$.
As mentioned above, regular maintenance of the various GLT systems, including the cryostat, could not be carried out during the pandemic, leading to a slight degradation of the performance of the SIS detector, which needs to be cooled to $\sim4$~K at 230 GHz. When we attempted to measure the efficiency again after the EHT 2021 observations, the sensitivity had deteriorated further, and we did not obtain any useful efficiency data at 230\,GHz. We therefore only have the antenna efficiencies measured a year before in 2020 as reference.

On the other hand, since the 86 GHz receiver is a MMIC amplifier that does not require 4~K temperatures, its sensitivity is not expected to be affected by the cryostat issues. We obtained 7 measurements of $\eta_{\rm A}$ at 86 GHz on Nov.10, 2020 toward Mars between elevations of $10^{\circ}-18^{\circ}$, obtaining $\eta_{\rm A}$ values of $75.3\pm0.9\%$ and $74.8\pm1.7\%$ for RCP and LCP, respectively. As with the measurements at 230\,GHz, these values are consistent with the antenna surface accuracy of $40 \, \mu$m rms, assuming the geometrical antenna efficiency of $75\%$. This suggests that the surface accuracy did not change significantly over the summer between March 2020 and November 2020. We argue that it is reasonable to assume that the 230\,GHz antenna efficiency measurements from March 2020 can be applied to the 2021 EHT observations.

The DPFU for the GLT for the EHT 2021 observations is thus estimated to be $0.0275 \pm 0.0015$\,K/Jy for RCP and $0.0265 \pm 0.0015$\,K/Jy for LCP. The uncertainties include the 5\% uncertainties of the flux density model for Venus added in quadrature.

\section{Note on Elevation and Time-Dependent Gains}\label{gaincurve}

To date, variations of the GLT antenna efficiencies as a function of elevation have not yet been determined to model the gain curves. However, due to the high latitude location of the GLT, astronomical sources do not significantly change in elevation as they move across the sky. Therefore, the elevation-dependent gains can be assumed to be insignificant, compared to the other uncertainties in the antenna efficiency measurements. Although there is a lack of sources that the GLT can track over a large range of elevations to measure the gains, we are working to measure the antenna efficiencies of the GLT as a function of time of day, ambient temperature, and elevation (where possible) to better characterize the telescope gains for future EHT campaigns. 

\section{Pointing Uncertainties}\label{pointing}

Radio pointing models for the GLT are determined from observations of bright spectral line (SiO $J=2-1$ masers at 86\,GHz, CO $J=2-1$ at 230\,GHz, and CO $J=3-2$ at 345\,GHz) sources, mostly evolved stars with rather compact and symmetrical emitting regions. To determine pointing offsets, the integrated intensity of a source spectral line is measured at 5 different points: at the expected source center, and at $\pm X \arcsec$ in azimuth and elevation, where $X$ is typically selected to be the half-width at half maximum (HWHM) of the primary beam, i.e. $15''$ at 230\,GHz, or larger if the source size is more extended than the beam HWHM. We then fit a Gaussian function to the line intensities in elevation and in azimuth to estimate the pointing offsets at these two axes. After obtaining a large sample of offset measurements for a large range of antenna elevations and azimuths, we fit the pointing model to the measured offsets as a function of telescope elevation and azimuth, from which we derive the pointing model coefficients \citep[see][for details on the pointing model]{koayetal20,chenetal23}.

\subsection{Pointing Uncertainties During the 2018 Observations}\label{pointing2018}

Prior to the 2018 April EHT campaign, pointing measurements were carried out at the GLT between March 16 and 22. Due to the low antenna efficiency at that time (see Sect.~\ref{Aeff2018}), we were limited to very bright sources as pointing sources, which were IRC$+$10216, CIT\,6, NGC\,7027, CRL\,2688, CRL\,618, and R\,Cas. Using about 70 pointing measurements at different azimuths and elevations, we updated the pointing model, then carried out more observations to verify the new model, finding an rms pointing accuracy of $\lesssim 4 \arcsec$ in both azimuth and elevation, for elevations above $10^{\circ}$. The low antenna sensitivity meant that pointing sources were observed at very low SNR at elevations below $10^{\circ}$, thereby increasing the uncertainties of the pointing offsets.

Just before the EHT 2018 observations, we found pointing elevation offsets of $10\arcsec$ at elevations of $10^{\circ}$, increasing to $40\arcsec$ offsets at elevations of $6^{\circ}$. By then we did not have time to conduct more pointing observations to correct the pointing model. This significantly impacts observations of 3C\,279, as it was observed between an elevation of 6$^\circ$ to 8$^\circ$ at the GLT, but should not affect observations of M87* (observed between $18^\circ$ to $26^\circ$ in elevation). 

%The pointing model was continually checked on the days leading up to the EHT2018 campaign to verify its accuracy.

The 230\,GHz FWHM beam sizes were measured to be 25\farcs5 in the horizontal (azimuth) direction, and 36\farcs7 in the vertical (elevation) direction, based on vertical and horizontal scans of the Moon. The beam elongation in the vertical direction relative to the horizontal direction is likely caused by the saddle shape of the antenna surface. Assuming a Gaussian-shaped beam, the rms pointing errors thus translate to amplitude uncertainties of $\lesssim 7$\% for sources observed at elevations above $10^{\circ}$. At low elevations, the pointing errors can cause the amplitudes to drop by 18\% at $10^{\circ}$ elevation, up to as much as 96\% at $6^{\circ}$. 

\subsection{Pointing Uncertainties in 2021}\label{pointing2021}

With the improved sensitivities of the antenna, we have been able to better characterize the pointing model of the GLT down to an rms accuracy of $\sim 2\arcsec$ at elevations as low as $7^{\circ}$ at 230\,GHz. Between 8 and 18 March 2021, just before the EHT campaign, we obtained 100 pointing measurements over a large range of azimuths and elevations. In addition to the bright pointing sources already used for 2018, we have added new (weaker) evolved star sources, such as $\chi$\,Cyg, GL\,3068, TX\,Cam, LP\,And, V384\,Per, IRAS\,21282+5050, GX\,Mon, V\,Cyg, RX\,Boo, and HD\,235858 to our catalogue of pointing sources.

Using the updated beam sizes after the panel adjustments of 27\farcs0, in both the horizontal and vertical axes, this translates to amplitude losses of $\lesssim 2$\% for all azimuths and elevations down to $7^{\circ}$ for the 2021 EHT observations. We expect to be able to achieve this pointing accuracy for subsequent campaigns.

\section{Summary}\label{summary}

In Table~\ref{2018vs2021}, we summarize the GLT performance parameters during the 2018 and 2021 EHT observing campaigns, as described in this memo.

\begin{table*}[ht]
\caption{Summary and comparisons of the GLT performance between 2018 and 2021.}\label{2018vs2021}
\begin{tabularx}{\linewidth}{p{2.5cm} p{6.5cm} p{6.5cm}}
\hline\hline       
 & \multicolumn{1}{c}{2018} & \multicolumn{1}{c}{2021} \\ 
\hline
\\
$T_{\rm sys}^{*}$ & Due to continuum detector instabilities, $T_{\rm sys}^{*}$ values derived based on the nominal receiver temperature, plus measured atmospheric temperatures and line-of-sight opacities for each scan. & $T_{\rm sys}^{*}$ values measured via the standard method of cold sky and hot load voltages, using the more stable outputs of the power-meter. \\
\\
$\eta_{\rm A}$\textsuperscript{a} & $21.6 \pm0.9\%$ for RCP and LCP & $67.1 \pm1.3\%$ for RCP \\
& & $64.7 \pm2.0\%$ for LCP \\
\\
DPFU\textsuperscript{b} & $0.0088 \pm 0.0006$\,K/Jy for both RCP and LCP & $0.0275 \pm 0.0015$\,K/Jy for RCP \\
& & $0.0265 \pm 0.0015$\,K/Jy for LCP\\
\\
Pointing losses &  $\lesssim 7$\% amplitude losses for elevations $> 10^{\circ}$;  &  $\lesssim 2$\% amplitude losses for elevations $> 7^{\circ}$ \\ 
 & $18\%$ to $96$\% amplitude losses between elevations of 10 to $6^{\circ}$ & \\
 \\
\hline
\end{tabularx}
\newline
\textsuperscript{a}\footnotesize{The $\eta_{\rm A}$ uncertainties quoted here comprise only the measurement errors.}\newline
\textsuperscript{b}\footnotesize{The DPFU uncertainties quoted here include the measurement uncertainties and the 5\% uncertainty of the Venus flux density model.}\newline
\end{table*}

\section*{Acknowledgement}

The authors would like to thank Claus Gonnsen, Ane-Louise Ivik, Chen-Yu Yu, Lupin C. Lin, Zheng Meyer-Zhao, Alex Alladi, Aaron Faber, and Nicolas Pradel for their expertise and assistance to the development of the GLT project. The GLT is supported by the Academia Sinica (Taiwan), the National Science and Technology Council (formerly the Ministry of Science and Technology) of Taiwan (grants 99-2119-M-001-002-MY4, 103-2119-M-001-010-MY2, 105-2112-M-001-025-MY3, 105-2119-M-001-042, 106-2112-M-001-011, 106-2119-M-001-013, 106-2119-M-001-027, 106-2923-M-001-005, 107-2119-M-001-017, 107-2119-M-001-020, 107-2119-M-001-041, 107-2119-M-110-005, 107-2923-M-001-009, 108-2112-M-001-048, 108-2112-M-001-051, 108-2923-M-001-002, 109-2112-M-001-025, 109-2124-M-001-005, 109-2923-M-001-001, 110-2112-M-003-007-MY2, 110-2112-M-001-033, 110-2124-M-001-007, 110-2923-M-001-001, 111-2124-M-001-005 and 112-2124-M-001-014) and the Smithsonian Institution. Support for the Hydrogen Maser frequency standard used for VLBI at the Greenland Telescope was provided through an award to SAO from the Gordon and Betty Moore Foundation (GBMF-5278). The GLT project members thank the National Chun-Shan Institute of Science and Technology (Taiwan) for their strong support, the National Astronomical Observatory of Japan for their support in receiver instrumentation, the United States National Science Foundation Office of Polar Programs for effective support in logistics, and the United States Air Force, 821st Air Base Group, Pituffik Air Base, Greenland for use of the site and access to the base and logistics chain. The GLT project members appreciate the development work done by Vertex Antennentechnik GmbH and ADS international, and also thank Atunas, an outdoor wear company in Taiwan, for sponsoring the Arctic wear used by staff. NRAO and the Massachusetts Institute of Technology Haystack Observatory supported the acquisition of the ALMA–North America prototype antenna and its re-purposing for deployment in Greenland. The GLT project acknowledges the privilege of operating on the lands of Greenland and strives to work with the local government and the people of Greenland to realize the scientific and educational potential of this project.

\end{document}